\documentclass[preprintnumbers,superscriptaddress,showkeys,showpacs,byrevtex]{revtex4}

\usepackage{amsmath,amsfonts,amssymb,amscd,amsxtra,amsthm}
\usepackage{graphicx}
\usepackage{epstopdf}
\usepackage{bm}
\usepackage{feynmp}
\usepackage{feynmp-auto}
\usepackage{amsmath,amsfonts,amssymb,amscd,amsxtra,amsthm}
\usepackage[colorlinks=true, pdfstartview=FitV, bookmarks=true, bookmarksnumbered=true, breaklinks]{hyperref}
\usepackage{physics}
\usepackage{multirow}
\usepackage[colorlinks=true, pdfstartview=FitV, bookmarks=true, bookmarksnumbered=true, breaklinks]{hyperref}
\usepackage{mathtools,braket}
\usepackage{soul}
\usepackage{lipsum} 
\usepackage{slashed}
\usepackage{xcolor}
\usepackage{physics}
\usepackage{multirow}

\definecolor{blue}{rgb}{0.0, 0.0, 1.0}
\definecolor{red}{rgb}{1.0, 0.0, 0.0}
\definecolor{royalblue}{rgb}{0.0, 0.14, 0.4}
\hypersetup{linkcolor=royalblue, citecolor=blue, urlcolor=royalblue}

\usepackage{hyperref}
\hypersetup{colorlinks=true,citecolor=blue,linkcolor=blue,urlcolor=blue}

\usepackage[mathlines]{lineno}

\usepackage{tikz,xcolor,hyperref}

\definecolor{lime}{HTML}{A6CE39}
\DeclareRobustCommand{\orcidicon}{%
	\begin{tikzpicture}
	\draw[lime, fill=lime] (0,0) 
	circle [radius=0.16] 
	node[white] {{\fontfamily{qag}\selectfont \tiny ID}};
	\draw[white, fill=white] (-0.0625,0.095) 
	circle [radius=0.007];
	\end{tikzpicture}
	\hspace{-2mm}
}

\foreach \x in {A, ..., Z}{%
	\expandafter\xdef\csname orcid\x\endcsname{\noexpand\href{https://orcid.org/\csname orcidauthor\x\endcsname}{\noexpand\orcidicon}}
}

\begin{document}
\preprint{PKNU-NuHaTh-2024, ADP-24-03/T1242}
\title{Neutrino mean free path in neutron stars in the presence of hyperons}

\author{Jesper Leong}
\email[E-mail: ]{jesper.leong@adelaide.edu.au}
\affiliation{CSSM and ARC Centre of Excellence for Dark Matter Particle Physics, Department of Physics, University of Adelaide, SA 5005 Australia}

\author{Parada T.~P.~Hutauruk\orcidB{}}
\email[E-mail: ]{phutauruk@pknu.ac.kr, phutauruk@gmail.com}
\affiliation{Department of Physics, Pukyong National University (PKNU), Busan 48513, Korea}

\author{Anthony W.~Thomas\orcidC{}}
\email[E-mail: ]{anthony.thomas@adelaide.edu.au}
\affiliation{CSSM and ARC Centre of Excellence for Dark Matter Particle Physics, Department of Physics, University of Adelaide, SA 5005 Australia}

\date{\today}

\begin{abstract}
We investigate the neutrino elastic differential cross-section (NDCS) and corresponding mean free path for neutral current scattering in the dense matter of a neutron star. A wide range of observed neutron star (NS) masses is considered,  including the presence of $\Lambda$, $\Xi^{-}$, and $\Xi^{0}$ hyperons in the heaviest stars. 
Their presence significantly decreases the total neutrino mean free path in the heavier stars.
\end{abstract}

\keywords{neutrino mean free path, neutral current, dense nuclear matter, neutron star, quark-meson coupling model, hyperons}
\maketitle
\section{Introduction}
During the evolution process of binary neutron star (NS) mergers~\cite{LIGOScientific:2017ync,LIGOScientific:2017vwq,Friman:1979ecl} and supernova collapse~\cite{Brown:1982cpw,Burrows:1987zz} neutrino emission plays a crucial role. Neutrinos are one of the group of multi-messengers, which along with gravitational waves~\cite{LIGOScientific:2017ync,LIGOScientific:2017vwq,LIGOScientific:2016aoc}, electromagnetic radiation~\cite{LIGOScientific:2017ync} and X-ray bursts~\cite{LIGOScientific:2017ync} offer vital information on the physical processes at play. Neutrino emission is expected to provide new insights into NS properties, with the information obtained able to be used to constrain the equation of state (EoS) of nuclear matter at extreme densities. The latter is still poorly determined and raises many open questions. The neutrinos produced in the process of NS mergers interact with the constituent matter of the NS through the weak interaction. This may involve either neutral current (NC) or charged current (CC) scattering. In this work, we will concentrate on neutrino NC scattering, which is particularly important for muon neutrinos~\cite{Bollig:2017lki}.

Thus far a number of studies of neutrino NC scattering and CC absorption in compact stars have been made using either relativistic~\cite{Reddy:1997yr,Hutauruk:2021cgi,Hutauruk:2023nqc} or non-relativistic models~\cite{Reddy:1997yr,Hutauruk:2023nqc,Mornas:2004vt}. However, most available calculations for the neutrino scattering or absorption have only considered matter consisting of protons and neutrons, with very few calculations considering the existence of the $\Lambda$ hyperon in the neutrino mean free path (NMFP) calculation~\cite{Mornas:2004vt,Mornas:2004vs,Reddy:1996tw,Rios:2006ic}. On the other hand, the vast differences in the time scales for weak interactions compared with NS formation time, along with considerations of $\beta$-equilibrium, lead one to expect that, even at zero temperature, hyperons must appear as stable constituents of the dense matter as the central baryon density increases and the NS become more massive. Of course, at the high temperatures experienced in the few milliseconds after a merger, hyperons will be abundant~\cite{Stone:2019blq,Oertel:2016xsn,Sekiguchi:2011mc,Panda:2010zz,Sedrakian:2022ata}.

Given the expected appearance of the $\Lambda$, $\Xi^{-}$ and even $\Xi^{0}$ hyperons at higher baryon number density, it is necessary to consider the interactions of the neutrinos with these hyperons in the NS. Here we compute the neutrino interactions with nucleons and include the $\Lambda$, $\Xi^{-}$, and $\Xi^{0}$ hyperons, which naturally appear in the high density EoS within the quark-meson coupling (QMC) model. It is both interesting and important to understand their contribution to the total NMFP. In this context, it is worth noting that the NMFP is required as input for supernova simulations~\cite{Fiorillo:2023frv,Liebendoerfer:2003es}.

In this paper, we compute the neutrino differential cross-section (NDCS) and NMFP for neutrinos scattering through the NC from baryons within NS of various masses. This necessarily includes neutrino-neutron, neutrino-proton, neutrino-$\Lambda$, neutrino-$\Xi^{-}$, and neutrino-$\Xi^{0}$ elastic scatterings. We do not consider the neutrino-electron and neutrino-muon scattering in the present work, as their contributions to the NMFP are small compared to those for the baryons~\cite{Niembro:2001hd,Hutauruk:2018cgu}. 

In order to calculate the NDCS and NMFP, we model the NS nuclear matter using the QMC model~\cite{Guichon:1987jp,Guichon:1995ue,Martinez:2018xep,Stone:2016qmi,Saito:2005rv,Saito:1994ki,Leong:2023lmw,Whittenbury:2015ziz}. This is a relativistic model which takes into account the modification of the internal quark structure of the baryons in-medium in response to the strong scalar mean field present. The $\sigma$, $\delta$, $\omega$, $\rho$, and $\pi$ mesons carry the interactions between baryons by coupling to the quarks confined in MIT bags. While pions are involved only through Fock terms, the $\sigma$, $\delta$, and the time components of the $\omega$ and $\rho$ fields constitute mean fields. Taking account of the change in the internal structure of the baryons as a result of the scalar mean-field is equivalent to the introduction of repulsive many-body forces~\cite{Guichon:2004xg,Guichon:2006er} between all the baryons, nucleons {\em and} hyperons. Because the meson interactions are defined at the quark level, no additional parameters are associated with either the hyperon-meson couplings or the many-body forces; they are all calculated in the model.

At the high densities of the nuclear matter associated with the heaviest NS, Pauli effects arising from the overlap of the finite-size baryons may lead to additional repulsion, beyond that experienced at normal nuclear matter density. We follow Leong {\em et al.}~\cite{Leong:2023yma} in treating this phenomenologically, in a manner that ensures that the properties of symmetric nuclear matter are unchanged at saturation density. With this additional repulsion, the model generates NS including hyperons with masses as large as 2.2 M$_\odot$.

Our computations reveal interesting predictions for the NDCS and NMFP arising from neutrino-$\Lambda$ and  neutrino-$\Xi$ scattering at higher baryon number densities. 
That $\Sigma$ hyperons do not occur in dense matter in $\beta$-equilibrium in the QMC model has been explained in earlier 
work~\cite{Guichon:2018uew}. The reason
is that the gluonic hyperfine interaction, which splits the masses of the $\Lambda$ and $\Sigma$ hyperons in free space, is enhanced in-medium as the mean scalar field increases~\cite{Guichon:2008zz}, raising their chemical potential so they cannot appear.
The NDCS for neutrino-$\Xi^{-}$ hyperon scattering begins to dominate beyond $n_B \simeq$ 4.0 $n_0$.
This increase in NDCS leads to a decrease in the NMFP at higher densities and therefore in heavier NS.

This paper is organized as follows. Section~\ref{sec:matterqmc} briefly describes the modeling of nuclear matter and the resultant EoS in the QMC model, including the phenomenological overlap correction.
The effective nucleon, hyperon masses, and particle fractions for different NS masses, serve as inputs to the neutrino-nucleon and neutrino-hyperon scattering calculations for the corresponding NS masses. In Sec.~\ref{sec:neuint}, we present the expressions for the NDCS and NMFP for the neutrino-nucleon and neutrino-hyperon interactions using a linear response theory approach at zero temperature. In Sec.~\ref{sec:nurests}, we present and discuss our numerical results for the NDCS and NMFP for different NS masses. Section~\ref{sec:sumarry} is devoted to a summary and concluding remarks. 

\section{Equation of State of Nuclear Matter}
\label{sec:matterqmc}
Here we use the QMC model to calculate the properties of nuclear matter in $\beta$-equilibrium. The model originated with Guichon~\cite{Guichon:1987jp}, who considered the effect of the strong Lorentz scalar field in a dense medium on the internal quark structure of the nucleons. Using the MIT bag model to describe nucleon structure he showed that the self-consistent treatment of the change in quark structure led to a novel saturation mechanism. Later refinements of the model involved a number of technical improvements as well as a generalization to include hyperons~\cite{Tsushima:1997cu,Guichon:2008zz}. 


In the QMC model, the coupling constants of the meson fields to the valence quarks are chosen such that a self-consistent calculation reproduces the properties of nuclear matter at normal saturation density. The model has been widely and successfully applied to many problems in nuclear physics, such as finite nuclei~\cite{Martinez:2018xep} and hadron structure in a nuclear medium~\cite{Saito:2005rv} as well as neutron star properties and possible hybrid stars~\cite{Whittenbury:2015ziz}. 
Here, we briefly review the EoS of nuclear matter within the QMC model, including the additional phenomenological repulsion which may arise at higher densities because of baryon overlap~\cite{Leong:2023yma}. 
Within this model, the baryon energy density takes the form
\begin{eqnarray}
    \label{eq1}
    \epsilon_B &=& \frac{\langle \mathcal{H}_B \rangle + \langle \mathcal{V}_\sigma \rangle + \langle \mathcal{V}_\omega \rangle + \langle \mathcal{V}_\rho \rangle + \langle \mathcal{V}_\delta \rangle  + \langle \mathcal{V}_\pi \rangle + \langle \mathcal{H}_O \rangle }{V} \, ,
\end{eqnarray}
where the baryon contribution in the first term of 
Eq.~(\ref{eq1}) is expressed as
\begin{eqnarray}
\label{eq1a}
\frac{\langle \mathcal{H}_B \rangle}{V} &=& 2 \sum_f \int^{k_f}_0  \frac{d^3k}{(2\pi)^3} \sqrt{\vec{k}^2 + M_f^{*2}}
\end{eqnarray}
and the meson contributions, including Fock terms, are 
\begin{eqnarray}
    \label{eq2}
    \frac{\langle \mathcal{V}_\sigma \rangle}{V} &=& \frac{m_\sigma^2 \bar{\sigma}^2}{2} + \frac{\lambda_3 g_\sigma^3 \bar{\sigma}^3}{6} + \sum_f \left( \frac{\partial M_f^*}{\partial \bar{\sigma}}\right)^2 \int^{k_f}_0 \int^{k_f}_0 \frac{d^3k_1 d^3k_2}{(2\pi)^6} \frac{1}{(\vec{k}_1-\vec{k}_2)^2 + m_\sigma^2} 
    \frac{M_f^{*2}} {\sqrt{(\vec{k}_1^2 + M_f^{*2})(\vec{k}_2^2 + M_f^{*2})}}, \\
    \frac{\langle \mathcal{V}_\omega\rangle}{V} &=& \frac{m_\omega^2 \bar{\omega}^2}{2} - \sum_f g_\omega^{f 2} \int^{k_f}_0 \int^{k_f}_0 \frac{d^3k_1 d^3k_2}{(2\pi)^6} \frac{1}{(\vec{k}_1-\vec{k}_2)^2 + m_\omega^2},\\
    \frac{\langle \mathcal{V}_\rho \rangle}{V} &=& \frac{m_\rho^2 \bar{\rho}^2}{2} - \sum_{f,f'} g_\rho^2 \int^{k_f}_0 \int^{k'_f}_0 \frac{d^3 k_1 d^3 k_2}{(2\pi)^6} \frac{C_{m,m'} \delta_{S,S'}}{(\vec{k}_1-\vec{k}_2)^2 +m_\rho^2}, \\
    \frac{\langle \mathcal{V}_{\delta}\rangle}{V} &=& \frac{m_\delta^2 \bar{\delta}^2}{2} + \sum_{f,f'} \left(g_\delta^f (\bar{\sigma}) g_\delta^{f'} (\bar{\sigma})\right) \int^{k_f}_0 \int^{k'_f}_0 \frac{d^3k_1 d^3k_2}{(2\pi)^6} \frac{C_{m,m'} \delta_{S,S'}}{(\vec{k}_1 -\vec{k}_2)^2 + m_\delta^2} \frac{ M_f^* M_{f'}^*} {\sqrt{(\vec{k}_1^2 + M_f^{*2})(\vec{k}_2^2+M_{f'}^{*2})}} \, .
\end{eqnarray}
Here the isoscalar and isovector-scalar~\cite{Motta:2019tjc} mean fields are denoted $\bar{\sigma}$ and $\bar{\delta}$, respectively. The $\bar{\omega}$ and $\bar{\rho}$ stand for the time components of the isoscalar and isovector-vector mean fields and $m_\sigma$, $m_\omega$, $m_\rho$, and $m_\delta$ are respectively $\sigma$, $\omega$, $\rho$, and $\delta$ meson masses. The $\sigma$ meson-nucleon, $\omega$ meson-nucleon, $\rho$ meson-nucleon, and $\delta$ meson-nucleon coupling constants in free space are respectively given by $g_\sigma$, $g_\omega$, $g_\rho$, and $g_\delta$. We follow Ref.~\cite{Leong:2023yma} in choosing $G_\delta \, (\equiv \, g_\delta^2/m_\delta^2) \, = 3$ fm$^2$, with $m_\delta = 983$ MeV. For the mass of the mesons $m_\sigma = 700$ MeV, with $m_\omega$ and $m_\rho$ taking their physical values. Finally $\lambda_3 = 0.02$ fm$^{-1}$, which sets the strength of the $\sigma$ self interaction term. The couplings of the $\sigma$, $\omega$, and $\rho$ were then fixed to reproduce the binding energy per nucleon $E_B/A =-$ 15.8 MeV and symmetry energy $S = 30$ MeV at the saturation density $n_0 =$ 0.16 fm$^{-3}$. The incompressibility was calculated as $K_\infty=260$ MeV \cite{Leong:2023yma}. The quantity $C_{m,m'} = \delta_{m,m'} I_m^{f2} + (\delta_{m,m'+1} + \delta_{m+1,m'}) I_t^f$, where the subscripts $m$ and $t$ stand for the projection, and isospin, respectively. 

The effective mass $M_f^* $ for each flavor baryon, $f$, which is particularly important in the present context, is given by
\begin{eqnarray}
\label{eq3}
 M_f^{*} \left( \bar{\sigma},\bar{\delta} \right) &=&  M_f - g_\sigma^f (\bar{\sigma}) \bar{\sigma} - g_\delta^f (\bar{\sigma}) I_{m}^f \bar{\delta}, \nonumber \\
 &=& M_f - \omega_\sigma^f g_\sigma \bar{\sigma} + \tilde{\omega}_\sigma^f \frac{d}{2} (g_\sigma \bar{\sigma})^{2}-t_\delta^f g_\delta I_m^f \bar{\delta} + \tilde{d} g_\sigma g_\delta \bar{\sigma} I_m^f \bar{\delta} \, .
\end{eqnarray}
Here the scalar polarizabilities, $d$ and $\tilde{d}$ (with numerical values given in Ref.~\cite{Motta:2019tjc}), are the origin of the repulsive many-body forces, which appear naturally and are determined, within the quark model used, with no new parameters. 
The sixth term on the right-hand side of 
Eq.~(\ref{eq1}) is the Fock term for pion 
exchange~\cite{Rikovska-Stone:2006gml,Krein:1998vc}.

As explained earlier, following Ref.~\cite{Leong:2023yma}, we also include a purely phenomenological repulsive contribution to the energy density which increases as the density, and hence the degree of potential overlap of the baryons, increases.
This short-distance repulsive contribution was shown as the last term in Eq.~(\ref{eq1}) and has the form
\begin{eqnarray}
    \label{eq6}
    \frac{\langle \mathcal{H}_O \rangle}{V} &=& E_0~n_B \exp \Bigg[ -\left( \frac{n_B^{-\frac{1}{3}}}{b}\right)^2\Bigg].
\end{eqnarray}
Here $n_B^{-1/3}$ is a measure of the average distance between baryons in the nuclear matter. The quantities $E_0$ and $b$ are free parameters, chosen such that they do not change the properties of symmetric NM at nuclear saturation density. These were found to be $E_0=5500$ MeV and $b=0.5$ fm \cite{Leong:2023yma}. The same values are used with $E_0$ describing the strength of the interaction and $b$ interpreted as the range parameter, corresponding to the size of the quark cores within the baryons. 

In the NS, we impose the conditions of $\beta$ equilibrium and charge neutrality. Then, the total energy density can be written as
\begin{eqnarray}
    \label{eq7}
    \epsilon_{\rm{total}} &=& \epsilon_B + \epsilon_e + \epsilon_{\mu} \, , 
\end{eqnarray}
where $\epsilon_B$ is defined in Eq.~(\ref{eq1}), while for each species of lepton, the energy density is defined as $\epsilon_l = 2 \int^{k_F^l}_0 d^3k/(2\pi)^3 \sqrt{\vec{k}_l^2 + m_l^2}$, with $m_l$ the lepton mass in free space.
\begin{figure}[t]
    \centering
    \includegraphics[width=0.95\columnwidth]{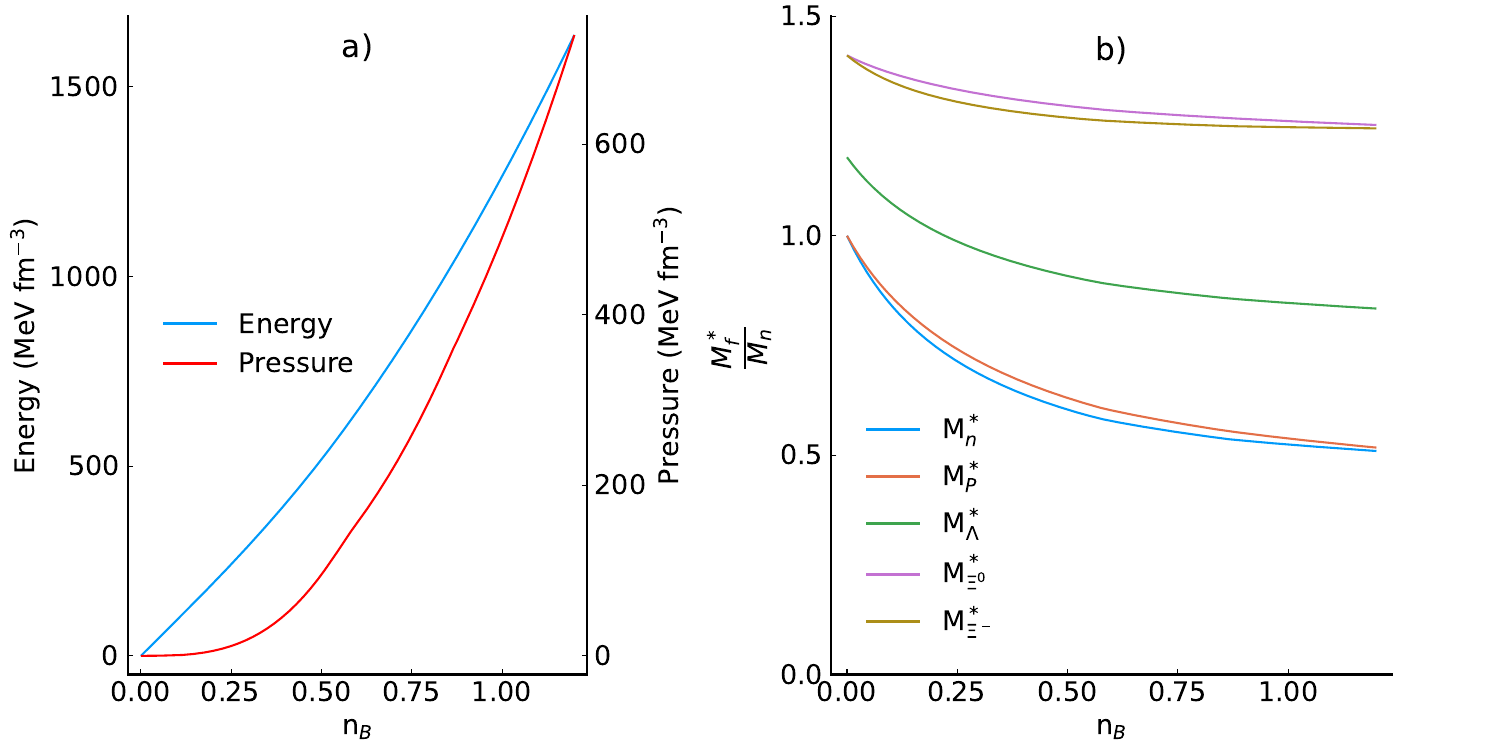}
    \caption{ (a) QMC EoS is displayed as a function of $n_B$ with the energy density (blue solid line) corresponding to the left vertical axis, and pressure (red solid line) corresponding to the right vertical axis. In (b), the effective mass, normalised to the free mass of the neutron ($M_n$), of each of the baryons falls in response to the increase in the mean scalar field strength (see Eq.~(\ref{eq3})).}
    \label{fig1}
\end{figure}
The pressure and energy density used here are shown as a function of the baryon number density in Fig.~\ref{fig1}(a), while the nucleon and hyperon effective masses are shown in Fig.~\ref{fig1}(b). It is worth noting that the EoS in Fig.~\ref{fig1}(a) was used to describe the NS properties starting from the most common masses (of order 1.5 $M_\odot$) up to the most massive stars~\cite{LIGOScientific:2018cki,Xie:2019sqb,Antoniadis:2013pzd,Ozel:2016oaf,NANOGrav:2017wvv,Riley:2021pdl,Fonseca:2021wxt,LIGOScientific:2020aai,Bassa:2017zpe}. 
 In Fig.~\ref{fig1}(b) we see that the effective masses, normalised to the free mass of the neutron, for different flavor baryons decrease as the baryon number density increases. In QMC, hyperons do not couple to the scalar field as strongly as the nucleons and as a result their effective mass does not decrease as much. $\Xi^{-,0}$ carries a strangeness number of 2 and thus its effective mass even at higher density does not greatly differ from it's true mass. The decrease is not as great as in many other models because of the self-consistent nature of the QMC model, with the internal structure of the baryons adjusting to oppose the applied scalar field.
 It is worth noting that in Fig.~\ref{fig1}, we only show the EoS and nucleon and hyperon effective masses over the range of baryon densities relevant to stable NS. 
\begin{figure}[t]
    \centering
    \includegraphics[scale=1]{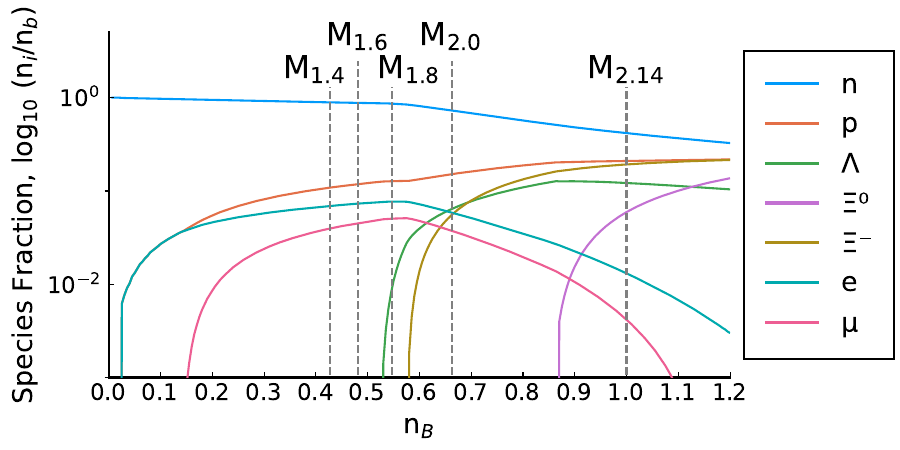}
    \caption{ Particle fractions for matter in $\beta$-equilibrium within the NS are calculated using the QMC model over the range $0<n_B<1.2$ fm$^{-3}$. Vertical dashed lines from left to right indicate a NS with a mass of 1.4, 1.6, 1.8, 2.0, and 2.14 $M_\odot$ respectively, with the latter the maximum predicted by the QMC EoS. All particle species falling to the left of the respective mass star are present within that star.}
    \label{fig2}
\end{figure}

Besides the EoS and nucleon and hyperon effective masses, we also compute the particle fractions for different NS masses. Results for these particle fractions for the $M_{\rm{NS}} =$ 1.4 $M_\odot$, which is the canonical NS mass~\cite{LIGOScientific:2018cki} are shown in Fig.~\ref{fig2}.
Of course, protons, neutrons, electrons, and muons are well-known as the standard matter occurring in NS. None of the hyperons appear in the low mass NS. This can be understood in terms of the maximum baryon number density for $M_{\rm{NS}} =$ 1.4 $M_\odot$, which is only $n_B \sim$ 0.43 fm$^{-3}$. A similar indication is found for $M_{\rm{NS}} =$ 1.6 $M_\odot$, with the same particles appearing. For $M_{\rm{NS}} =$ 1.8 $M_\odot$ the $\Lambda$ hyperon starts to appear at threshold baryon number density of around $n_B =$ 0.55 fm$^{-3}$, as shown in Fig.~\ref{fig2}. 
As the baryon number density increases further the $\Xi^{-}$ hyperons begin to appear -- beyond the relevant baryon number density threshold, which lies around $n_B =$ 0.58 fm$^{-3}$. The threshold density for the $\Xi^0$ is greater still at $n_B=0.87$ fm$^{-3}$. The QMC model predicts a maximum mass neutron star ($M_{\rm{max}}=2.14 \, M_\odot $) to have a central core density in excess of 6 $n_0$, which at its centre, is the highest density achievable. In consequence, the composition of heavy neutron stars, such as PSR J0740+6620 with $M_{\rm{NS}}= 2.072^{+0.067}_{-0.066}$ $M_\odot$, is predicted by the QMC model to have the $\Lambda$, $\Xi^-$, and $\Xi^0$ hyperons present within the inner core. Finally, once each of the hyperons appears, they continue to increase in abundance as the baryon number density grows (or the mass of the NS increases), which is likely to affect the NDCS and NMFP. For heavy NS the model predicts that, in abundances, the dominant hyperonic species is the $\Xi^-$ followed by the $\Lambda$ and finally the $\Xi^0$.  Note there are a wide variety of relativistic mean field models which also vary in their hyperonic content. Schaffner-Bielich~\cite{Schaffner-Bielich:2008zws} shows a similar ordering of hyperons as QMC, although the $\Lambda$ remains the most abundant of all the hyperons. In contrast Reddy \textit{et al.} \cite{Reddy:1997yr} showed that along with the $\Lambda$ the $\Sigma^{0,\pm}$ also appear whilst the $\Xi^{0,-}$ are absent.

\begin{figure}[t]
    \centering
    \includegraphics[width=0.90\columnwidth]{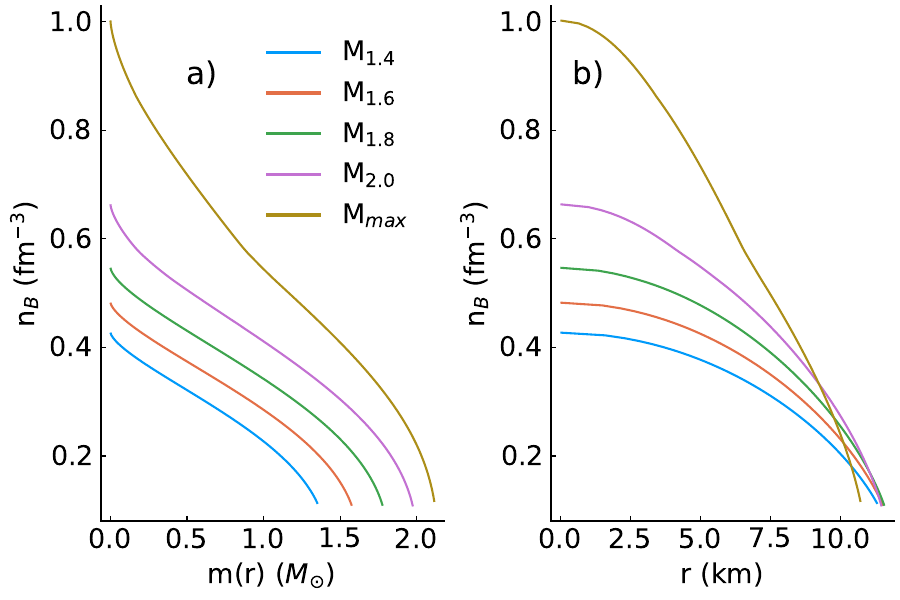}
    \caption{ (a) Baryon number density as a function of the $m(r)/M_{\odot}$ where $m(r)$ is the total energy within a certain radius in a star, (b) Baryon number density as a function of radius.}
    \label{fig3}
\end{figure}

As a further illustration of the physics of NS, in Fig.~\ref{fig3}, we show the baryon number density at various radii within the different chosen fixed mass NSs. In Fig.~\ref{fig3}(a), the baryon number density, $n_B$, is given at the mass contained within a given radius, $m(r)$ in units of $M_\odot$. Figure~\ref{fig3}(b) shows how the baryon number density changes from the centre as we move radially outwards towards the surface. In comparison to a canonical NS mass, $M_{\rm{NS}}=1.4$ $M_\odot$, the central density of our maximum mass NS is around 2.3 times larger. This corresponds to the change of the energy density and pressure of 2.9 and 8.3 times larger, respectively. Even when going from $M_{\rm{NS}}=2.0$ $M_\odot$ to the maximum $M_{\rm{NS}}=2.14$ $M_\odot$, the central density and consequently the energy density and pressure (see Fig.~\ref{fig1}(a)), need to increase by a large amount for relatively small gains in mass. This indicates that the highest mass NSs have extremely high pressure to counteract the gravitational force.

At this point, one can summarize that the underlying cold NS EoS used here is derived from the QMC model which produces NS masses up to and beyond 2.0 $M_\odot$. The central densities of NS can be in excess of 6 times the saturation density of symmetric nuclear matter. This is a density region where additional interactions may be present~\cite{Harvey:1980rva,HALQCD:2018gyl} but difficult to probe in laboratory experiments~\cite{Liu:1993sc}. 
At zero temperature the EoS used here predicts that only the $\Lambda$, $\Xi^0$, and $\Xi^-$ hyperons will appear in even the heaviest stars, while $\Sigma$ hyperon will not appear as previously explained.
Next, we use the nucleon and hyperon effective masses and particle fractions computed for different NS masses to calculate the NDCS and NMPF.

\section{Neutrino Interaction}
\label{sec:neuint}
In this section, we briefly introduce the general formalism for the neutrino interaction with NS matter through NC scattering using linear response theory. Earlier work involved the calculation of NDCS and NMFP, including weak~\cite{Hutauruk:2022bii} and electromagnetic interactions. That work took into account the neutrino form factors such as the neutrino magnetic moment and charge radius~\cite{Hutauruk:2022bso,Hutauruk:2020mhl,Hutauruk:2018cgu,Hutauruk:2006re,Kalempouw-Williams:2005zbp,Sulaksono:2006eu}. However, to avoid complications, in the present work we limit our investigation to the Standard Model weak interaction. The following theoretical framework follows the work presented in Ref.~\cite{Hutauruk:2022bii} (see also \cite{Reddy:1997yr} and \cite{Horowitz:1990it}). The relevant interaction Lagrangian for the neutrino-baryon and neutrino-hyperon NC scatterings in terms of the current-current interaction is written as
\begin{eqnarray}
\label{eqscat1}
    \mathcal{L}_{\mathrm{INT}}^{NC} &=& \frac{G_F}{\sqrt{2}} \Big[ \bar{\nu}_e \gamma^\mu \left( 1-\gamma_5 \right) \nu_e \Big] \Big[ \bar{\psi}_{B,Y} \Gamma_\mu^{[B,Y],NC} \psi_{B,Y} \Big] \, ,
\end{eqnarray}
where $G_F = 1.023/M \times 10^{-5}$ is the weak coupling constant and for standard nucleon vertex (including free space form factor of a nucleon) is defined $\Gamma_\mu^{[B,Y],NC} = \gamma_\mu \Big[C_V^{[B,Y]} - C_A^{[B,Y]} \gamma_5 \Big]$, where $B = p,n$ and $Y=\Lambda, \Xi^{-}$, and $\Xi^{0}$ hyperons. The values of $C_V$ and $C_A$ for neutrons, protons, and $\Lambda$, $\Xi^{-}$, and $\Xi^{0}$ hyperons can be found in Refs.~\cite{Hutauruk:2022bii,Savage:1996zd,Carrillo-Serrano:2014zta}. They are summarized in Table.~\ref{tab2}. 
\begin{table}[h]
    \centering
  \caption{Axial and vector couplings for proton, neutron, and $\Lambda$, $\Xi^{-}$, and $\Xi^{0}$ hyperons at $q^2=0$ for neutral-current reactions. In the numerical calculation, we use $\sin^2\theta_w = 0.231$, $g_A=1.27$, $\kappa_p=1.793$ and $\kappa_n=-1.913$. The values of $D = 0.774$ and $F = 0.496$ are considered in this work. 
  }
\label{tab2}    
  \begin{tabular}{*{3}{p{4.5cm}}}\hline \hline
    Target & $C_A$ & $C_V$ \\ \hline
    $n$ &$-(D+F)/2 = -\frac{g_A}{2}=-0.635$  & $-0.5$ \\
    $p$ & $ (D+F)/2 = \frac{g_A}{2}=0.635$ & $0.5 - 2 \sin^2 \theta_w =0.038$  \\
    $\Lambda$ & $-F/2-D/6 = -0.377$ & $-0.5$  \\
     $\Xi^{-}$ & $ (D-3F)/2=-0.357$ & $-\frac{3}{2} + 2 \sin^2 \theta_w= -1.038$  \\
      $\Xi^{0}$ & $-(D+F)=\frac{-g_A}{2}=-0.635$ & $-0.5$  \\
 \hline
  \end{tabular}
\end{table}

An expression for the double differential cross-section per volume for the neutrino scattering can be computed from the interaction Lagrangian given in Eq.~(\ref{eqscat1}) and it gives
\begin{eqnarray}
\frac{1}{V} \frac{d^3 \sigma}{d^2 \Omega' dE_{\nu}'} = - \frac{G_F^2}{32\pi^2} \frac{E_{\nu}'}{E_{\nu}} \mathrm{Im} \Big[L_{\mu \nu} \Pi^{\mu \nu}_{[B,Y]} \Big] \, , 
\end{eqnarray}  
where $E_\nu$ and $E'_\nu = E_\nu -q_0$ are the initial and final neutrino energies. The leptonic and hadronic tensors are respectively defined by
\begin{eqnarray}
    L_{\mu \nu} = 8\Big[ 2 k_\mu k_\nu + (k\cdot q) g_{\mu \nu} - (k_\mu q_\nu + q_\mu k_\nu) \mp i \epsilon_{\mu \nu \alpha \beta k^\alpha q^\beta} \Big] \, ,
\end{eqnarray}
where the sign in the last term of the leptonic tensor is minus ($-$) for neutrinos and plus ($+$) for antineutrinos.
In addition, we need
\begin{eqnarray}
    \Pi_{\mu \nu}^{[B,Y]} (q^2) = -i \int \frac{d^4 p}{(2\pi)^4} \mathrm{Tr} \Big[ G^{[B,Y]} (p) \Gamma_\mu^{[B,Y]} G^{[B,Y]} (p+q) \Gamma_\nu^{[B,Y]} \Big] \, , 
\end{eqnarray}
where the nucleon and hyperon propagators in the nuclear medium are defined by
\begin{eqnarray}
    G^{[B,Y]} (p) = \Bigg[ \frac{p\!\!\!/^{*}+M_f^*}{p^{*2} - M_f^{*2} + i\epsilon} + i\pi \frac{p\!\!\!/^{*} + M_f^{*}}{E^{*}} \delta \left( p_0^* -E^*\right) \Theta \left(p_F^{[B,Y]} - |\mathbf{p}| \right)\Bigg] \, ,
\end{eqnarray}
where $M_f^*$ is the nucleon or hyperon effective mass as defined in Eq.(\ref{eq3}) and $p_{F}$ being the appropriate Fermi momentum.

After contracting the leptonic tensor and the polarization insertions (hadronic and hyperonic tensors) for the neutrons, protons, and hyperons, the final expression for the NDCS is given by
\begin{eqnarray}
   \frac{1}{V} \frac{d^3\sigma}{dE_\nu^{'} d^2 \Omega'} = \frac{G_F^2}{4\pi^3} \frac{E_\nu^{'}}{E_\nu} q^2 \Bigg[ A \mathcal{R}_1 + \mathcal{R}_2 + B \mathcal{R}_3 \Bigg] \, . 
\end{eqnarray}
Here $A = \left[ 2E_\nu (E_\nu -q_0) + 0.5 q^2\right]/|\vec{\mathbf{q}}|^2$, $B= 2E_\nu -q_0$, and $\mathcal{R}_1$, $\mathcal{R}_2$, and $\mathcal{R}_3$ are,  respectively, given by
\begin{eqnarray}
    \mathcal{R}_1 &=& \left( C_V^2 + C_A^2 \right) \Big[ \mathrm{Im} \Pi_L^{[B,Y]} + \mathrm{Im} \Pi_T^{[B,Y]} \Big], \nonumber \\
    \mathcal{R}_2 &=& C_V^2 \Pi_T^{[B,Y]} + C_A^2 \Big[ \mathrm{Im} \Pi_T^{[B,Y]} - \mathrm{Im} \Pi_A^{[B,Y]} \Big], \nonumber \\
    \mathcal{R}_3 &=& \pm 2 C_V C_A \mathrm{Im} \Pi_{VA}^{[B,Y]} \, .
\end{eqnarray}
The plus ($+$) sign in $\mathcal{R}_3$ is for the neutrinos and the minus ($-$) sign is for the antineutrinos. In a nuclear medium, the polarization insertion can be decomposed into the polarization for the longitudinal, transversal, axial, and mixed vector-axial channels for the neutrons, protons, and hyperons. These are, respectively, given by
\begin{eqnarray}
    \mathrm{Im} \Pi_L^{[B,Y]} &=& \frac{q^2}{2\pi |\vec{\mathbf{q}}|^3} \left[\frac{q^2}{4} \left( E_F-E^* \right) + \frac{q_0}{2} \left( E_F^2 -E^{*2}\right) + \frac{1}{3} \left( E_F^3 - E^{*3}\right)\right],\\
    \mathrm{Im} \Pi_T^{[B,Y]} &=& \frac{1}{4\pi |\vec{\mathbf{q}}|} \Bigg[ \left( M_f^{*2} + \frac{q^4}{4|\vec{\mathbf{q}}|^2} + \frac{q^2}{2} \right) \left( E_F -E^*\right) + \frac{q_0 q^2}{2|\vec{\mathbf{q}}|^2} \left( E_F^2 -E^{*2}\right) + \frac{q^2}{3|\vec{\mathbf{q}}|^2} \left( E_F^3 - E^{*3} \right) \Bigg], \\
    \mathrm{Im} \Pi_A^{[B,Y]} &=& \frac{i}{2\pi |\vec{\mathbf{q}}|} M_f^{*2} \left( E_F - E^{*} \right), \label{eq:polarA}\\
    \mathrm{Im} \Pi_{VA}^{[B,Y]} &=& \frac{iq^2}{8\pi |\vec{\mathbf{q}}|^3} \left[ \left( E_F^2 - E^{*2} \right) + q_0 \left( E_F-E^*\right)\right] \, .
\end{eqnarray}

The final inverse NMFP expression for the neutrino-nucleon and neutrino-hyperon NC scatterings as a function of the initial energy of neutrino at zero temperature is given by
\begin{eqnarray}
\label{eq:nmfp}
    \lambda^{-1} (E_\nu) = 2\pi \int_{q_0}^{(2E_\nu -q_0)} d|\vec{\mathbf{q}}| \int_0^{2E_\nu} dq_0 \frac{|\vec{\mathbf{q}}|}{E_\nu E_\nu'} \left[ \frac{1}{V} \frac{d^3 \sigma}{dE_\nu' d^2\Omega'}\right] \, .
\end{eqnarray}
It is worth noting that in the calculation of the total neutrino mean free path, we first calculate the neutrino cross-section for each particle of the system using the linear response approximation. With the cross-sections, we then calculate the mean free path using Eq.~(\ref{eq:nmfp}) for particles of the system. The explicit expression for the total mean free path can be written
\begin{eqnarray}
    \label{eq:totalNMFP}
    \lambda^{-1}_{\textrm{total}} &=& \lambda^{-1}_p + \lambda^{-1}_n + \lambda^{-1}_{\Lambda} + \lambda^{-1}_{\Xi^0} + \lambda^{-1}_{\Xi^-}, 
\end{eqnarray}
where the subscripts of $p$, $n$, $\Lambda$, $\Xi^0$, and $\Xi^-$ are respectively proton, neutron, $\Lambda$, $\Xi^0$, and $\Xi^-$.

\section{Numerical results and Discussions}
\label{sec:nurests}
\begin{figure}[t]
    \centering
    \includegraphics[width=0.9\columnwidth]{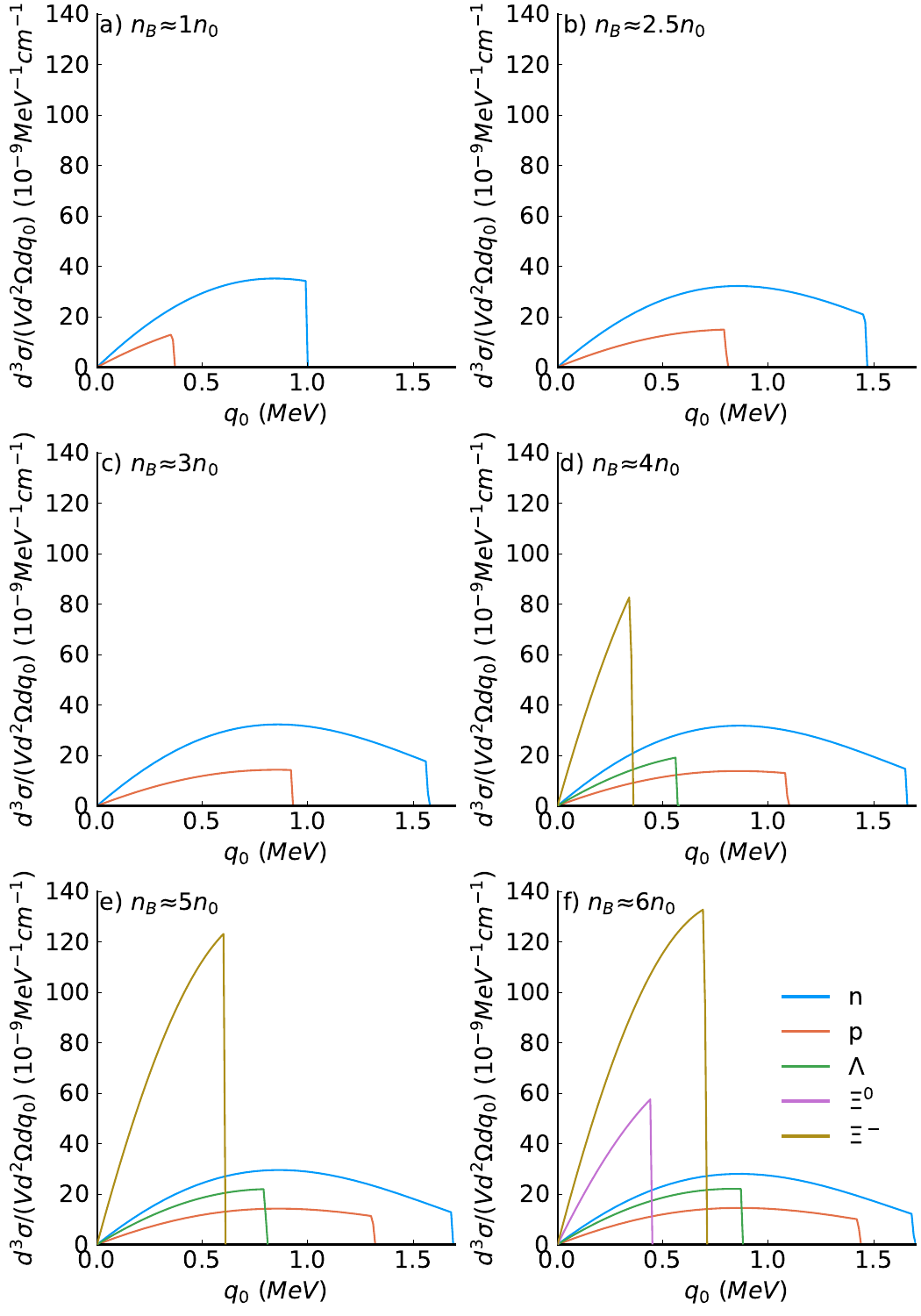}
    \caption{ Neutrino differential cross section was calculated using $E_\nu=5$ MeV and $|\vec{\mathbf{q}}|=2.5$ MeV, with the corresponding baryon number density given from (a) through to (f).}
    \label{fig8}
\end{figure}

\begin{figure}[t]
    \centering
    \includegraphics[width=0.9\columnwidth]{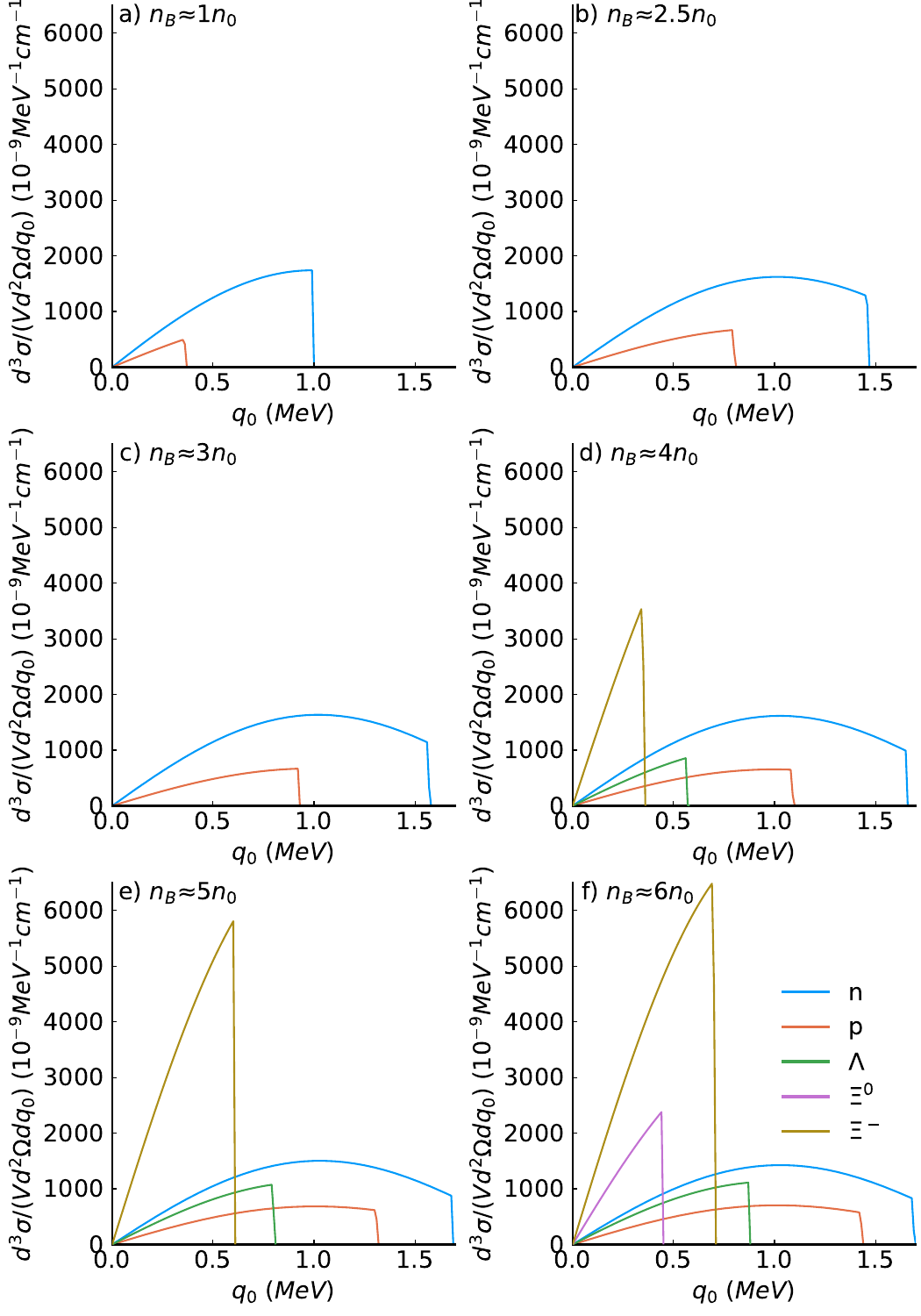}
    \caption{ Same as in Fig.~\ref{fig8}, but for $E_\nu =$ 30 MeV.}
    \label{fig:dcs30}
\end{figure}

\begin{figure}
    \centering
    \includegraphics{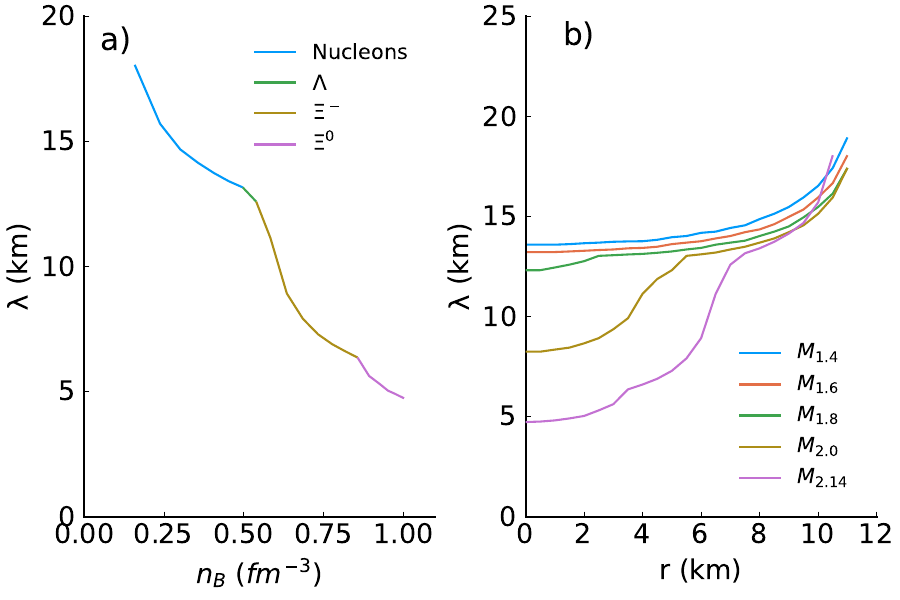}
    \caption{(a) Neutrino mean free path for $E_\nu=5$ MeV as a function of the baryon number density. The color scheme reflects the appearance of new baryon species as the number density increases. In (b) Neutrino mean free path as a function of NS radius $r$ for stars of various masses, from 1.4 to 2.14 $M_\odot$. }
    \label{fig:MFP5}
\end{figure}

\begin{figure}
    \centering
    \includegraphics{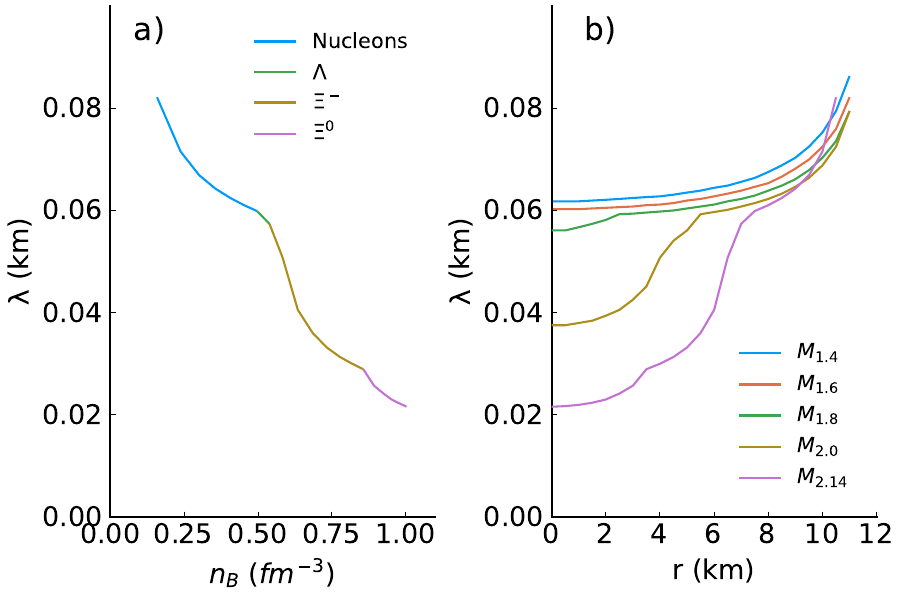}
    \caption{Same as in Fig.~\ref{fig:MFP5}, but for $E_\nu =$ 30 MeV.}
    \label{fig:MFP5R}
\end{figure}

The numerical results for the NDCS and NMFP of neutrino-nucleon and neutrino-hyperon scattering for various NS masses, computed in the QMC model including the repulsive baryon overlap term, are presented in Figs.~\ref{fig8}-\ref{fig:dcs30} and in Figs.~\ref{fig:MFP5}-\ref{fig:MFP5R}, respectively. The NDCS and NMFP are first computed at three-component momentum transfer $|\vec{\textbf{q}}| =$ 2.5 MeV and initial neutrino energy $E_\nu =$ 5 MeV (Figs. \ref{fig8} and \ref{fig:MFP5}), which is typical of the neutrino energies in NS during the cooling phase~\cite{Horowitz:1990it}. 
In Figs.~\ref{fig:dcs30} and \ref{fig:MFP5R} we also show results of relevance to supernovae and NS mergers for the commonly used neutrino energy around $E_\nu =$ 30 MeV~\cite{Janka:2006fh,Kresse:2020nto,Bruenn:1985en,Kamiokande-II:1987idp}.

 As depicted in Fig.~\ref{fig2}, the central densities of stars with mass in the range 1.4$-$1.6 $M_\odot$ are around $2.5-3.0$ $n_0$. Thus we compute the NDCS starting at very low baryon number density $n_B \simeq$ 1.0 $n_0$ (the crust-core boundary of the NS), then at $n_B \simeq$ 2.5 $n_0$, which is the central core density of a $M_{\mathrm{NS}} =$ 1.4 $M_\odot$, and finally at $n_B \simeq$ 3.0 $n_0$, which represents the central density of a $1.6$ $M_\odot$ NS. 

The NDCS results for $M_{\rm{NS}} =$ 1.4 $ M_\odot$ are shown in Figs. $\ref{fig8}$(a)-(b). For the canonical mass NS case, we find that only the proton and neutron appear (see Fig.~\ref{fig2}) and, as a consequence, only the nucleons contribute to the NDCS. As shown in Fig.~\ref{fig2}, the electrons and muons also appear together with protons and neutrons but their NDCS are relatively small compared to those for protons and neutrons and we ignore them in the present calculation. As expected, the NDCS for the neutrons is greater than that of the protons for all baryon number densities. This can be understood because of the differences in the abundances of the particles present within the star, where the neutron particle fractions are larger in comparison to those for the protons. 

The sharp peak structure in NDCS, shown in Figs.~\ref{fig8} and \ref{fig:dcs30} depends on the maximum value of $q_0$, $q_0^{\rm{max}}\approx|\vec{\mathbf{q}}|/\sqrt{(M_f^*/p_F)^2+1}$~\cite{Hutauruk:2022bii}, which increases as the effective mass decreases. 
Here we note that the appearance of protons and neutrons for 1.4 $M_\odot$ NS is consistent with the result found in Ref.~\cite{Zhao:2015tra}, except for the appearance of the $\Lambda$ hyperon in that work. That depends, of course, on the depth of the hyperon potential and coupling constant in the model. The QMC model provides a relatively good description of the binding energies of known hypernuclei, which are consistent with the available experimental results~\cite{Guichon:2008zz}. 

Results for the NDCS at densities relevant to a star with $M_{\rm{NS}} =$ 1.8 $M_\odot$ are shown in Figs.~\ref{fig8}(a)-(c), where the baryon number density varies from $n_B \simeq$ 1.0 and 2.0 to 3.0 $n_0$. In Fig.~\ref{fig8} we do not show explicitly the NDCS result for $n_B \simeq$ 3.4 $n_0$, where the $\Lambda$ hyperons start to appear. At this density, which is just reached in a 1.8 $M_\odot$ star, the NDCS for neutrino-$\Lambda$ hyperon scattering does begin to contribute to the total NDCS. However, as the number density is still very low, the NDCS for the $\Lambda$ hyperon for $M_{\rm{NS}} =$ 1.8 $M_\odot$ is very small. The range of $q_0$ is also small compared to that for the neutrons at $n_B \simeq$ 3.4 $n_0$. 

For $M_{\rm{NS}} =$ 2.0 $M_\odot$, not only does the neutrino-$\Lambda$ hyperon NDCS contribute but the neutrino-$\Xi^{-}$ hyperon also starts to contribute to the total. The contributions at $n_B \simeq$ 4.0 $n_0$ can be seen in Fig.~\ref{fig8}(d). This is consistent with the threshold baryon number density for the appearance of the $\Xi^{-}$ hyperon as shown in Fig.~\ref{fig2}. 
Figures~\ref{fig8}(e) and (f) show the NDCS relevant to a star with mass $M_{\rm{NS}} = M_{\rm{max}}$, where the maximum baryon number density rises just beyond $n_B \simeq$ 6.0 $n_0$. Here all 3 hyperons, $\Lambda$, $\Xi^-$, and $\Xi^0$, contribute to the NDCS, although unevenly. The $\Xi^-$ hyperon has the greatest influence on the total NDCS which can be explained in terms of the relative abundances and respective effective masses. From Fig.~\ref{fig2}, $\Xi^-$ makes up a higher proportion of a heavy NS core when compared to the $\Lambda$ and, subsequently, the $\Xi^0$ hyperon. It's abundance is roughly equal to the nucleons however it's contribution to the NDCS is several times greater than the neutron (see Fig. \ref{fig2}). This is a consequence of the axial polarisation insertion in Eq. \ref{eq:polarA}, which is governed by the size of $M^*$ \cite{Hutauruk:2022bii}. Fig. \ref{fig2} shows that the effective mass of the $\Xi^{-,0}$ is the largest of all the baryons and thus the axial channel is increased leading to a larger NDCS. This explains the differences shown in the NDCS between the baryons even if their abundances are closely matched.

%

As explained earlier, we also compute the NDCS for higher neutrino energy, $E_\nu =$ 30 MeV, which is relevant to supernova and NS mergers, as given in Fig.~\ref{fig:dcs30}. We find that the NDCS for $E_\nu =$ 30 MeV has the same tendency and interpretation as $E_\nu =$ 5 MeV for appropriate baryon number densities, except for the order of magnitude. The NDCS increases remarkably, by a factor of 50, at higher nuclear matter density, leading to a much lower NMFP, as shown in Fig.~\ref{fig:MFP5R}(a). 

Next we turn to the neutrino mean free path. Results for the total NMFP for $E_\nu =$ 5 MeV as a function of $n_B$ for $M_{\rm{NS}}= M_{\rm{max}}$ are shown in Fig.~\ref{fig:MFP5}(a), covering a large range of baryon number density. Here we consider only the total NMFP at $M_{\rm{NS}}= M_{\rm{max}}$, treating it as representative of other NS, since the same EoS is used to describe them.
At low density the only contribution to the NMFP comes from the nucleons (as indicated by the blue line). This occurs up to around $n_B\simeq 0.54$ fm$^{-3}$ where a small plateau precedes a slight but rapid decrease in the NMFP caused by the appearance of the $\Lambda$ hyperon (green). There is a second more significant decrease to the NMFP which closely follows the $\Lambda$ and is associated with the $\Xi^-$ appearance at around $n_B\simeq 0.67$ fm$^{-3}$ (mustard). Finally for the heaviest NS the central core density is in excess of 0.83 fm$^{-3}$ and the $\Xi^0$ hyperon contributions further decreases the NMFP. 

In Fig.~\ref{fig:MFP5} (b), we show the NMFP for $E_\nu=$ 5 MeV as a function of radius for different NS masses. We see that the NMFP results for $M_{\rm{NS}} =$ 1.4 M$_\odot$ and 1.6 M$_\odot$ are almost the same because the contributions to the total NMFP in both cases come from the protons and neutrons. For $M_{\rm{NS}} =$ 1.8 M$_\odot$, the $\Lambda$ hyperon begins to contribute to the total NMFP, in addition to the protons and neutrons, leading to a decrease in the total NMFP around the core of the NS, $r \simeq$ 2 km ($n_B \simeq $ 0.54 fm$^{-3}$, see Fig.~\ref{fig2}). This indicates that the neutrino emission could be delayed around the NS core. 

The decreases in NMFP begin to move to larger radii in the heavier stars. 
For example, for $M_{\rm{NS}} =$ 2.0 M$_\odot$ the NMFP starts to decrease rapidly inside about 6 km, where the $\Lambda$ hyperons appear. A little further inside, at around $r \simeq$ 5.5 km, the $\Xi^{-}$ begin to contribute. With the contribution of the $\Xi^0$ to the total NMFP for $M_{\rm{NS}} =$ 2.14 M$_\odot$, one sees a greater decrease in NMFP at around $r \simeq$ 7.8 km. From these results, one can conclude that mostly $\lambda > R_{\rm{NS}}(=r)$, indicating the neutrinos can escape from the NS relatively easily, leading to fast NS cooling. This also indicates that our results suggest very little chance of neutrino trapping for any cool NS. 

Next we compute the NMFP for $E_\nu =$ 30 MeV, showing the results in Figs.~\ref{fig:MFP5R}(a) and~\ref{fig:MFP5R}(b). The qualitative trend of the NMFP in Figs.\ref{fig:MFP5R}(a) and ~\ref{fig:MFP5R}(b) was found to be the same as that found in Figs.~\ref{fig:MFP5}(a) and~\ref{fig:MFP5}(b), respectively. However, they are very different in magnitude, as small as some tens of metres. Indeed, we found that the $\lambda << R_{\rm{NS}}$ for all NS masses, which is in contrast with the results in Fig.~\ref{fig:MFP5}(b). This indicates that neutrinos have a large probability of trapping in a hot NS environment.

\section{Summary and future perspectives}
\label{sec:sumarry}
To summarize, we have investigated the neutrino differential cross section (NDCS) and mean free path (NMFP) for neutrino-nucleon and neutrino-hyperon neutral current scattering at densities relevant to a wide range of neutron star masses. The QMC model was used to generate the equation of state for matter in $\beta$-equilibrium. This was then used in the TOV equations to calculate the structure of the NS. The relativistic effective masses of the nucleons and hyperons were then used in the NDCS and NMFP calculations, leading to predictions for the NMFP as a function of density and equivalently radius in the NS. 

For NDCS results for different NS masses, we found interesting results for the NDCS of the neutrino-$\Lambda$, neutrino-$\Xi^{-}$ and neutrino-$\Xi^{0}$ scattering in the matter with higher baryon density ($n_B \gtrsim$ 3.4 $n_0$), corresponding to NS masses beyond $M_{\rm{NS}} =$ 1.8 $M_\odot$. Within the QMC model with additional repulsion depending on the degree of overlap of the baryons, we found that the threshold baryon number densities for $\Lambda$, $\Xi^{-}$, and $\Xi^{0}$ hyperons are around 0.53 fm$^{-3}$, 0.58 fm$^{-3}$, and 0.87 fm$^{-3}$, respectively. The NDCS for the $\Xi^{-}$ begins to dominate beyond $n_B \simeq$ 4.0 $n_0$.  

An increase in the NDCS of neutrino-$\Xi^{-}$ hyperons at higher baryon number density implies that the NMFP decreases. As the $\Xi^{0}$ hyperon appears at a very high threshold baryon number density, around $n_B \simeq$ 6.0 $n_0$, the NDCS of the neutrino-$\Xi^{0}$ hyperon scattering starts to contribute to the total NDCS at that density.

Next, we found that the NMFP for a star at the canonical NS mass, $M_{\rm{NS}} =$ 1.4 $M_\odot$, decreases as density increases up to $n_B/n_0 \simeq$ 2.5. A similar behavior is shown by the total NMFP for $M_{\rm{NS}} =$ 1.6 $M_\odot$. Note that, in this range of NS masses, only the NMFP of protons and neutrons contribute to the total NMFP. This is because the hyperons do not yet appear at the baryon number densities corresponding to either NS mass. 
As the threshold baryon number density of the $\Lambda$ hyperon is reached, the NMFP of the neutrino-hyperon scattering for $M_{\rm{NS}} =$ 1.8 $M_\odot$ starts to contribute to the total NMFP, in addition to the NMFP of the neutrino-nucleon scattering. The NMFP contribution of the neutrino scattering by a particular hyperon is signaled by a further drop in the total NMFP just beyond the relevant threshold baryon number density. There are several such thresholds shown in Figs.~\ref{fig:MFP5} and \ref{fig:MFP5R} for $M_{\rm{NS}} =$ 2.0 $M_\odot$ and $M_{\rm{NS}} = M_{\rm{max}}$.

An important feature of the results presented here (in Fig.~\ref{fig:MFP5} versus Fig.~\ref{fig:MFP5R}) is that the NMFP shows a dramatic decrease as the neutrino energy goes up. In particular, a neutrino of energy 30 MeV has a mean free path of just 20 m in the core of the most heavy NS. As hyperons exhibit no threshold at finite temperatures and are more abundant at even 10 MeV than at zero temperature~\cite{Stone:2019blq,Guichon:2023iev}, the pressure associated with neutrino trapping by hyperons will be especially important in proto-NS.

\section*{Acknowledgment}
The authors wish to thank the Asia Pacific Center for Theoretical Physics (APCTP) for the support and warm hospitality and for organizing the 2023 workshop on \textit{Origin of Matter and Masses in the Universe: Hadrons in Free Space, Dense Nuclear Medium, and Compact Stars}, where the discussions on the topic of this paper were initiated. P.T.P.H was supported by the National Research Foundation of Korea (NRF) grants funded by the Korean government (MSIT) (2018R1A5A1025563, 2022R1A2C1003964, and 2022K2A9A1A0609176). We also acknowledge the support of the University of Adelaide and the Australian Research Council through Discovery Project (AWT) DP230101791.


\end{document}